\documentclass[pra,twocolumn,showpacs]{revtex4}%
\usepackage{graphicx}
\usepackage{amsmath}
\usepackage{amsfonts}
\usepackage{amssymb}
\usepackage{hyperref}
\usepackage{appendix}%
\providecommand{\U}[1]{\protect\rule{.1in}{.1in}}
\ifx\pdftexversion\undefined
\else
\fi
\expandafter\ifx\csname package@font\endcsname\relax\else
\expandafter\expandafter
\expandafter
\expandafter
\expandafter{\csname package@font\endcsname}
\fi
\begin{document}
\title{Manipulation of a Bose-Einstein condensate by a time-averaged orbiting
potential using phase jumps of the rotating field}
\date{\today}
\author{P.~W.~Cleary}
\author{T.~W.~Hijmans}
\author{J.~T.~M.~Walraven}
\affiliation{Van der Waals-Zeeman Institute of the University of Amsterdam,
Valckenierstraat 65, 1018 XE, The Netherlands}

\begin{abstract}
We report on the manipulation of the center-of-mass motion (`sloshing') of a
Bose Einstein condensate in a time-averaged orbiting potential (TOP) trap. We
start with a condensate at rest in the center of a static trapping potential.
When suddenly replacing the static trap with a TOP trap centered about the
same position, the condensate starts to slosh with an amplitude much larger
than the TOP micromotion. We show, both theoretically and experimentally, that
the direction of sloshing is related to the initial phase of the rotating
magnetic field of the TOP. We show further that the sloshing can be quenched
by applying a carefully timed and sized jump in the phase of the rotating field.

\end{abstract}

\pacs{37.10.Gh, 34.10.+x, 45.50.-j,03.75.Kk}
\maketitle

\section{Introduction}

Time averaged potentials (TAP) offer a versatile tool for trapping both
charged and neutral particles. For neutral atoms the most common example in
this class of traps is the Time-averaged Orbiting Potential (TOP) which was
used in the experiments in which the first Bose-Einstein condensate (BEC) was
created \cite{Cornell,BEC1995}. The TOP trap consists a magnetic quadrupole
trap \cite{Migdall85,Bergeman87} shifted by a uniform magnetic modulation
field rotating at a high (audio) frequency. As this rotation is slow as
compared to the Larmor precession of the atomic magnetic moments, the atoms
remain polarized with respect to the instantaneous effective magnetic field
\cite{Cornell98} as follows from the adiabatic theorem. On the other hand the
rotation is fast as compared to the orbital motion of the atoms. As a
consequence, the atomic motion consists of a fast rotating part (micromotion),
superimposed on a slow oscillating part (macromotion). In the simplest
theoretical description, the static approximation, the micromotion is
eliminated by time-averaging the instantaneous potential over a full cycle of
the modulation field.

Suppose we load a particle with given momentum $\mathbf{p}_{0}$ at position
$\mathbf{r}_{0}$ in a TOP trap using a sudden switch-on procedure. One might
naively guess that the ensuing motion is given by the dynamics in the
time-averaged potential, subject to the initial conditions\textbf{
}$\mathbf{r}=\mathbf{r}_{0}$ and $\mathbf{p}=\mathbf{p}_{0}$ but this guess
turns out to be wrong. In fact, one can show that the initial conditions for
the slow motion depend on the phase of the TOP at the time of switch-on. This
phenomenon was analyzed by Ridinger and coworkers \cite{Rid1,Rid2} for the
special case of a one-dimensional rapidly oscillating potential (ROP) with
zero average. Ridinger et al.~also showed, first for a classical particle
\cite{Rid1} and subsequently for the quantum case \cite{Rid2}, that the
amplitude and energy associated with the slow motion can be altered by
applying a suitable phase jump in the rapidly oscillating field.

In this paper we show, both theoretically and experimentally, that the
dependence on initial phase and the possibility to influence the motion by
phase jumps, is also present for a two-dimensional rotating TOP field. In
particular we show that a cloud of atoms which is initially at rest with zero
momentum acquires a sloshing motion as soon as the TOP is suddenly switched
on. This is true even if the cloud is initially at the minimum of the
effective potential. The amplitude of this slow macromotion is much larger
than that of the fast micromotion while the direction of sloshing depends on
the TOP phase at switch-on. We also demonstrate that this macromotion can be
almost entirely quenched by applying a carefully timed and sized phase jump in
the TOP field.

The motion of atoms and ultracold atomic clouds in TOP traps have been
extensively described in the literature. Following the achievement of the
first BEC \cite{BEC1995}, the use of the axially symmetric TOP was described
theoretically in \cite{Edwards,Yukalov97,Kuklov97,Minogin,Franzosi,Challis}
and explored experimentally by other groups
\cite{Heinzen,Phillips1,Kasevich,Arimondo,Foot1,Scherer} to study properties
of the BEC. The idea of the TOP was extended to an asymmetric triaxial TOP
trap developed by \cite{Phillips1} and also used by other groups
\cite{Arimondo, Scherer}. Further a number of other variations were
introduced: In many cases, it turns out to be convenient to switch on the TOP
after a preparative stage of cooling in a conventional static trap such as a
magnetic quadrupole trap (see e.g.\thinspace\cite{Phillips1}), an optically
plugged magnetic quadrupole \cite{Raman} and Ioffe-configurations
\cite{Tiecke03,Thomas,Buggle04}. Often, the transfer of the cloud from the
static to the TOP trap cannot be performed adiabatically for topological
reasons. Bearing this in mind, it becomes relevant to carefully analyze the
dynamics that may be induced by a sudden switch-on of the TOP. In addition,
applications which require manipulation of a BEC are heavily dependent on
precise control of the location of the atomic cloud and can thus benefit from
the techniques described.

In our experiments the condensate is prepared in a Ioffe-Pritchard (IP) trap
before transferring to a TOP because the use of radio-frequency (rf) induced
evaporative cooling is more efficient in a static magnetic trap, resulting in
larger condensates. Once transferred to the TOP we can create trapping
geometries that are difficult to realize using a static magnetic potential
without introducing Majorana losses associated with the presence of zero-field
points. An example is the double well potential used in \cite{Tiecke03}.

The remainder of this paper is organized as follows. In Section \ref{theory}
we calculate the motion of a cloud of atoms in a TOP which at switch-on is at
rest at the center of the trap. We discuss the motion that results and derive
the conditions under which a phase jump can lead to a substantial reduction of
the energy associated with the slow motion of the cloud. In Section
\ref{experimentaldetails} we discuss the experimental details and the
preparation of the BEC and its transfer to the TOP. In Section \ref{results}
we present the experimental results and compare with the theory of Section
\ref{theory}. Finally in Section \ref{Discussion and Outlook} we give a
summary and conclusion.

\section{Theory\label{theory}}

\subsection{Time-averaged Ioffe-Pritchard potential}

In the literature the term TOP is most often used for a spherical-quadrupole
trap combined with a rotating uniform magnetic modulation field. In this paper
we will use the term TOP in a broader context, to include the magnetic
trapping potential created by combining a IP trap with rotating modulation
field. Challis et al.~\cite{Challis} have shown that the dynamical eigenstates
of a degenerate Bose gas in a TOP are given by solutions of the usual
Gross-Pitaevskii equation but taken in a circularly translating reference
frame, that is, a reference frame the origin of which performs a rapid
circular motion but retains a constant orientation. In particular this implies
that the center of mass of a condensate in its ground state performs the same
micromotion in a TOP as a point particle with the magnetic moment of an atom.
In this spirit we use as a $^{87}$Rb condensate to study the micromotion and
macromotion in a TOP.

We consider a cigar-shaped Ioffe-Pritchard potential
\cite{LuitenThesis93,Bergeman87,Surkov94}
\begin{equation}
U(\boldsymbol{\varrho},z)=\mu\sqrt{\alpha^{2}\varrho^{2}+(B_{0}+\tfrac{1}%
{2}\beta z^{2})^{^{2}}}, \label{instantpotential}%
\end{equation}
where $\boldsymbol{\varrho}(t)$ is the radial position of a test atom with
respect to the IP symmetry axis, $\mu$ the magnetic moment of the atom, and
$\alpha$, $\beta$, $B_{0}$ the parameters for the radial gradient, the axial
curvature and offset value of the IP magnetic field. Eq.\thinspace
(\ref{instantpotential}) represents an approximate expression for the IP trap
which is valid for $\alpha^{2}\gg\beta B_{0}$ and in the limit $\varrho
\ll\alpha/\beta$ \cite{LuitenThesis93,Bergeman87,Surkov94}.

In the presence of the TOP field we transform to the circularly translating
frame \cite{Challis} and have%
\begin{equation}
\boldsymbol{\varrho}(t)=\{x-\rho_{m}\cos(\omega t+\phi_{m}),y-\rho_{m}%
\sin(\omega t+\phi_{m})\}, \label{transformation}%
\end{equation}
where $\{x,y,z\}\equiv\{\boldsymbol{\rho},z\}\equiv\mathbf{r}$ is the position
of the atom in the laboratory frame and the IP symmetry axis is displaced over
a distance $\rho_{m}=B_{m}/\alpha$ in the direction
\begin{equation}
\boldsymbol{\hat{\rho}}_{m}=\{\cos(\omega t+\phi_{m}),\sin(\omega t+\phi
_{m})\}
\end{equation}
by the uniform modulation field
\begin{equation}
\mathbf{B}_{m}=B_{m}\{\cos(\omega t+\phi_{m}),-\sin(\omega t+\phi_{m})\}
\end{equation}
applied perpendicular to the $z$ axis. The $y$ axis is taken along the
vertical direction, the $xz$ plane being horizontal. The modulation field
$\mathbf{B}_{m}$ rotates at angular frequency $-\omega$ (phase $-\phi_{m}$)
about the horizontal $z$ axis as illustrated in Fig.\thinspace
\ref{Fig:CircleOfDeath}. Notice that the sense of rotation of the
IP-field-minimum is opposite to that of the $\mathbf{B}_{m}$ field, in
contrast to the original TOP configuration \cite{Cornell}, where the
field-zero rotates in the same direction as the bias field. This reflects the
difference between the 2D-quadrupole symmetry of the IP trap and the axial
symmetry of the spherical-quadrupole trap. The rotation of the modulation
field $\mathbf{B}_{m}$ also gives rise to a fictitious field $\mathbf{B}%
_{\omega}$ which has to be added or subtracted from the offset field
$\mathbf{B}_{0}$, depending on the sense of rotation,
\begin{equation}
\mathbf{B}_{0}\rightarrow\mathbf{B}_{0}(1\pm\mathbf{B}_{\omega}/\mathbf{B}%
_{0})=\mathbf{B}_{0}(1\pm\omega/\omega_{L}),
\end{equation}
where $\omega_{L}=g_{F}\mu_{B}B_{0}/\hbar$ is the Larmor frequency of magnetic
moment of the atoms, with $g_{F}$ the hyperfine $g$ factor and $\mu_{B}$ the
Bohr magneton. In a standard TOP, the fictitious field in combination with
gradient of the quadrupole field gives rise to a shift of the equilibrium
position of the cloud in the direction of the axis around which the field
rotates \cite{Cornell98,Arimondo}. In our IP-TOP the axial field is
homogeneous near the origin and the shift is absent; the change in $B_{0}$
turns out to be small and will be neglected in this paper.

For $\beta=0$ and $B_{0}=0$ the potential $U(\boldsymbol{\varrho},z)$
corresponds to that of a two-dimensional quadrupole field with a zero-field
line that rotates at distance $\rho_{m}$ about the $z$ axis as a result of the
modulation. For $B_{0}=0$ the distance $\rho_{m}$ is known as the radius of
the `circle of death'. For $B_{0}<0$ the potential corresponds to two TOP
traps separated by $\Delta z=2(2|B_{0}|/\beta)^{1/2}$ \cite{Tiecke03}. In this
paper we will consider only the case $B_{0}\geq0$.

In the common description of the TOP one analyzes the motion in an effective
potential, obtained by time averaging the static trap over a full rotation
period of the $\mathbf{B}_{m}$ field. For Eq.\thinspace(\ref{instantpotential}%
) this procedure yields the effective potential%
\begin{equation}
\mathcal{U}(\mathbf{r})=\frac{1}{2\pi}\int_{0}^{2\pi}U(x-\rho_{m}\cos
\zeta,y-\rho_{m}\sin\zeta,z)d\zeta, \label{TOPpotential}%
\end{equation}
where $\zeta=\omega t+\phi_{m}$. For the cigar-shaped IP potential we consider
the condition
\begin{equation}
\omega\gg\Omega_{\rho}\gg\Omega_{z}, \label{topcondition}%
\end{equation}
where, for an atom of mass $m$, the quantity $\Omega_{z}=(\mu\ \beta/m)^{1/2}$
is the axial harmonic oscillation frequency in the effective potential
$\mathcal{U}(0,0,z)$. Analogously, harmonic oscillation frequency in the
radial plane is given by
\begin{equation}
\Omega_{\rho}=\sqrt{\frac{\mu\alpha^{2}}{m\bar{B}_{0}}(1-\tfrac{1}{2}B_{m}%
^{2}/\bar{B}_{0}^{2})}\equiv\Omega, \label{Omega-Big}%
\end{equation}
where $\bar{B}_{0}\mathbf{=(}B_{0}^{2}+B_{m}^{2})^{1/2}$ is offset value of
the effective potential at the origin \cite{Tiecke03}.

The first inequality in Eq.\thinspace(\ref{topcondition}) ensures that the
fast and slow radial motions of the atoms can be separated, which is the
well-known operating regime for a TOP trap \cite{Cornell}. The second
inequality implies that the axial motion in the effective trap is slowest and
that the motion can be treated as quasi two-dimensional in the radial plane.

To account for the acceleration due to gravity $\left(  g\right)  $, the
gravitational potential $mgy$ has to be added to Eqs.\thinspace
(\ref{instantpotential}) and (\ref{TOPpotential}). The main effect is to shift
the minimum of the potentials in the negative $y$ direction by the amount%
\begin{equation}
\Delta y=g/\Omega^{2}.
\end{equation}
This expression holds as long as the gravitational sag $\Delta y$ is much
smaller than the harmonic radius $\rho_{h}\equiv\bar{B}_{0}/\alpha$.

Since $\rho_{h}\geq\rho_{m}$, the effective potential (\ref{TOPpotential}) may
be treated as harmonic as long as the motion is confined to a region around
the $z$ axis that is small compared to $\rho_{m}$. For our experiment the
harmonic approximation holds rather well and is sufficient for gaining
qualitative insight in the micro- and macromotion as will be shown in Section
\ref{sec:micromotion and macromotion}. Refinements associated with switch-on
transients and gravity are discussed in Appendix
\ref{app:Delayed Sudden-Step Model}. In the numerical analysis of Section
\ref{sec:Numerical analysis}, we solve the classical equations of motion in
the full time-dependent potential Eq.\thinspace(\ref{instantpotential}). In
this context we also comment on the validity of the harmonic approximation.

\subsection{Micromotion and macromotion
\label{sec:micromotion and macromotion}}

\begin{figure}[ptb]%
\centering
\includegraphics[
trim=0.506072in 0.234148in 0.462478in 0.291894in,
height=7.5493cm,
width=7.8319cm
]%
{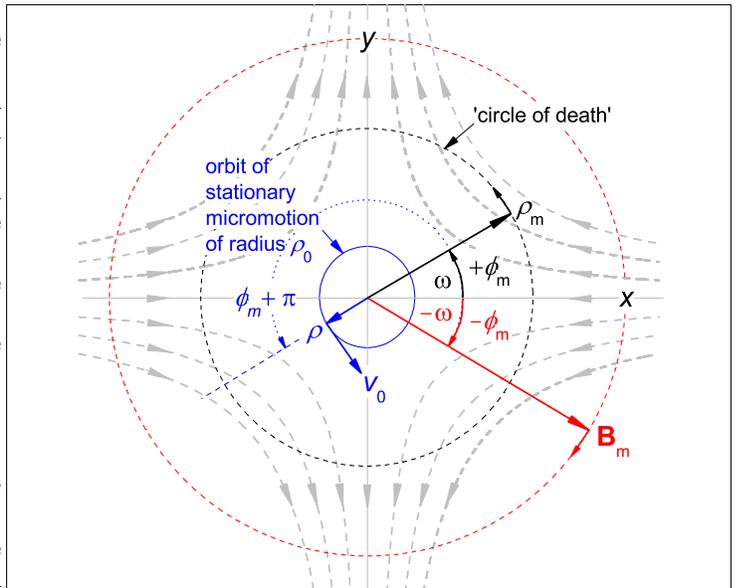}%
\caption{(color online) Schematic diagram of the magnetic field configuration
in relation to the orbit of stationary micromotion (solid blue circle). The
view is along the (horizontal) $z$ axis. The orbital position and velocity of
the micromotion are denoted by $\boldsymbol{\rho}=-\rho_{0}\boldsymbol{\hat
{\rho}}_{m}$ and $v_{0}$. The IP symmetry axis rotates at frequency $\omega$
(with initial phase $\phi_{m}$) about the $z$ axis on the circle of radius
$\rho_{m}$ (dashed black circle). Note that the TOP field $\mathbf{B}%
_{m}=B_{m}\boldsymbol{\hat{\rho}}_{m}$ rotates at frequency $-\omega$ (phase
$-\phi_{m}$), reflecting the 2D-quadrupole symmetry (dashed red circle) of the
IP trap.}%
\label{Fig:CircleOfDeath}%
\end{figure}

To analyze the effect of switching on the $\mathbf{B}_{m}$ field at $t=0$ we
first consider an atom `at rest' in the center of the effective trapping
potential $\mathcal{U}(\boldsymbol{\rho},z)$. Such an atom exhibits no
period-averaged dynamics (no macromotion) but only circular micromotion at a
frequency $\omega$ about the origin as illustrated in Fig.\thinspace
\ref{Fig:CircleOfDeath}. The radius of this stationary micromotion,%
\begin{equation}
\rho_{0}=\frac{\mu\alpha}{m\omega^{2}}\left(  1+B_{0}^{2}/B_{m}^{2}\right)
^{-1/2}, \label{rho-0}%
\end{equation}
follows from the condition $F_{c}=m\omega^{2}\rho_{0}$ for the centripetal
force $\mathbf{F}_{c}=-\boldsymbol{\nabla}_{\boldsymbol{\rho}}U|_{\rho=0}%
=\mu\alpha(1+B_{0}^{2}/B_{m}^{2})^{-1/2}\boldsymbol{\hat{\rho}}_{m}$. The
speed of this stationary micromotion,
\begin{equation}
v_{0}=\omega\rho_{0}=\frac{\mu\alpha}{m\omega}\left(  1+B_{0}^{2}/B_{m}%
^{2}\right)  ^{-1/2}, \label{vel-0}%
\end{equation}
is directed orthogonally to the direction $\boldsymbol{\hat{\rho}}_{m}$. Such
pure micromotion only results if at $t=0$ the atom is already moving at speed
$v_{0}$ along a circle of radius $\rho_{0}$ about the origin and is located at
position $\boldsymbol{\rho}=-\rho_{0}\boldsymbol{\hat{\rho}}_{m}$ (see
Fig.\thinspace\ref{Fig:CircleOfDeath}).%
Obviously an atom at $t=0$ at rest at the origin, $\boldsymbol{\rho}=\{0,0\}$
does not satisfy these initial conditions and as a consequence its macromotion
will start with a finite launch speed. We will see that the result is
elliptical motion at frequency $\Omega$, with the long axis approximately
perpendicular to the initial direction of $\boldsymbol{\hat{\rho}}_{m}$ and
with a substantial amplitude, of order $\left(  \omega/\Omega\right)  \rho
_{0}$. Usually this motion is undesired and our aim is to quantify it and
subsequently quench it by imparting a phase jump to the TOP-field.

To gain insight into the way in which the sudden switch-on of the TOP
influences the macromotion of an atom initially at rest at the origin, we
first consider a simple model in which it is assumed that the motion in the
radial plane can be decomposed into two harmonic components, oscillating at
the micromotion and macromotion frequencies $\omega$ and $\Omega$,
respectively. The position $\boldsymbol{\rho}(t)$ and velocity
$\boldsymbol{\dot{\rho}}(t)$ are given by%
\begin{align}
\boldsymbol{\rho}(t)  &  =\left\{  \rho_{0}\cos(\omega t+\phi),\rho_{0}%
\sin(\omega t+\phi)\right\}  +\nonumber\\
&  \ \ \ \ +\left\{  X_{0}\cos(\Omega t+\varphi_{x}),\ Y_{0}\sin(\Omega
t+\varphi_{y})\right\} \label{r_approx}\\
\boldsymbol{\dot{\rho}}(t)  &  =\left\{  -v_{0}\sin(\omega t+\phi),v_{0}%
\cos(\omega t+\phi)\right\}  +\nonumber\\
&  \ \ \ \ +\left\{  -V_{0,x}\sin(\Omega t+\varphi_{x}),\ V_{0,y}\cos(\Omega
t+\varphi_{y})\right\}  , \label{v_approx}%
\end{align}
where $X_{0}$ $(Y_{0})$ is the amplitude, $V_{0,x}=\Omega X_{0}$
$(V_{0,y}=\Omega Y_{0})$ the velocity amplitude and $\varphi_{x}$
$(\varphi_{y})$ the initial phase of the macromotion in $x$ $(y)$ direction;
$\phi$ is the initial phase of the micromotion. The atom starts at rest at the
origin, hence the initial conditions are $\boldsymbol{\rho},\boldsymbol{\dot
{\rho}}=0$ at $t=0$. If the condition
\begin{equation}
\omega\gg\Omega\label{TOPcondition}%
\end{equation}
is satisfied, the acceleration due to the micromotion dominates over that of
the macromotion. The total acceleration may be approximated by
$\boldsymbol{\ddot{\rho}}\simeq\mathbf{F}_{c}/m$. In other words,
$\boldsymbol{\ddot{\rho}}$ points in the direction $\boldsymbol{\hat{\rho}%
}_{m}$, which is opposite to the direction of $\boldsymbol{\rho}$ (as per
Fig.\thinspace\ref{Fig:CircleOfDeath}). Hence, the initial phase of the
micromotion is $\phi\simeq\phi_{m}+\pi$, where $\phi_{m}$ is fixed by the
phase of the rotating $\mathbf{B}_{m}$ field \cite{ApproxPhase}. Without loss
of generality we can set $\phi_{m}=0$, which means that $\boldsymbol{\hat
{\rho}}_{m}$ is oriented along the positive $x$ direction at $t=0$. With this
choice and setting $\phi=\phi_{m}+\pi$, we find from the initial conditions:
$\varphi_{x},\varphi_{y}=0$, $X_{0}=\rho_{0}$, and $Y_{0}=(\omega/\Omega
)\rho_{0}$. Substituting these values in Eq.\thinspace(\ref{r_approx}) we
obtain an equation for the macromotion representing an elliptical orbit with
its major axis oriented perpendicular to the instantaneous direction
$\boldsymbol{\hat{\rho}}_{m}$ of the $\mathbf{B}_{m}$ field at $t=0$. Since
the amplitude of the macromotion along its major axis is larger than the
micromotion by the factor $\omega/\Omega$, a substantial sloshing motion
results from the sudden switch-on. Note that with increasing $\omega$, the
micromotion amplitude $\rho_{0}$ decreases like $1/\omega^{2}$ whereas the
amplitude of the sloshing motion $Y_{0}$ decreases only like $1/\omega$. For
this reason the sloshing cannot be neglected in most practical cases involving
audio-frequency modulation.%
\begin{figure}[ptb]%
\centering
\includegraphics[
height=4.228cm,
width=8.2621cm
]%
{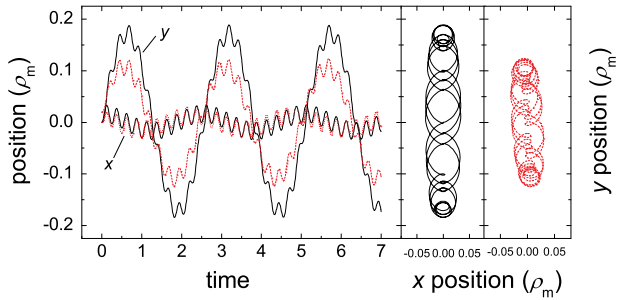}%
\caption{(color online) Numerically calculated trajectories in the $xy$ plane
with the $x$- and $y$-positions shown against time (left) and parametric plots
of the same trajectory in the $xy$ plane (middle and right) of a particle
initially at rest at the origin, after instant switch-on (black lines). The
dotted red curves correspond to a switch-on time of 3\textrm{~}$\mathrm{\mu
s}$\textrm{ }of the TOP field, a settling time for the value of $B_{0}$ as
well as the presence of gravity. The trap frequencies are $\omega/2\pi=4$ kHz
and $\Omega/2\pi=394~\mathrm{Hz}$. Units are scaled to the TOP radius
$\rho_{m}$.}%
\label{theortraject}%
\end{figure}

\subsection{Numerical analysis\label{sec:Numerical analysis}}

To validate the analytical model introduced in Section
\ref{sec:micromotion and macromotion}, we numerically integrate the classical
equations of motion in the full time-dependent potential given by
Eq.\thinspace(\ref{instantpotential}) for $z=0$, $v_{z}=0$ and $\phi_{m}=0$.
The result for the trajectory is given in Fig.\thinspace\ref{theortraject} and
exhibits the sloshing macromotion described above. The choice of parameters is
such that it matches the experimental conditions that will be presented in
Section \ref{experimentaldetails}.%

\begin{table}[t] \centering
\caption{Comparison of numerical results (num) with the analytical model (AM); +ab - including refinements (a) and (b); +abc - all refinements included}%
\begin{tabular}
[c]{c|c|c|c|c|c|c}\hline\hline
& $\phi_{m}$ & $\theta/\pi$ & $\varphi_{x}/\pi$ & $\varphi_{y}/\pi$ &
$X_{0}/\rho_{0}$ & $Y_{0}/\rho_{0}$\\\hline
{\small num} & $0$ & $0$ & $0$ & $0$ & $1$ & $10.2$\\
{\small AM} & $0$ & $0$ & $0$ & $0$ & $1$ & $10.2$\\\hline
{\small num+ab} & $0$ & $0.024$ & $0.22$ & $0.04$ & $1.34$ & $10.2$\\
{\small AM+ab} & $0$ & $0.024$ & $0.23$ & $0.04$ & $1.34$ & $10.2$\\\hline
{\small num+abc} & $0$ & $0.017$ & $0.20$ & $0.06$ & $0.82$ & $6.5$\\
{\small AM+abc} & 0 & $0.021$ & $0.23$ & $0.06$ & $0.85$ & $6.5$%
\end{tabular}
\label{table:FitResultsA}%
\vspace{3mm}
\end{table}%

The drawn black lines in Fig.\thinspace\ref{theortraject} correspond to sudden
switch-on of the TOP trap at $t=0$ for an atom initially at rest at the origin
in the absence of gravity. The figure clearly shows the micromotion
superimposed onto the macromotion orientated along the $y$ direction. The
amplitudes and phases of the macromotion obtained by fitting Eq.\thinspace
(\ref{r_approx}) to the results of the numerical calculation agree accurately
with the analytical model of Section \ref{sec:micromotion and macromotion}
(see Table~\ref{table:FitResultsA}). A more detailed comparison reveals that
anharmonicities play a minor role; the harmonics of both the micro- and
macromotion have amplitudes which are at least two orders of magnitude smaller
than those of the fundamentals.

In order to allow a better comparison with the experiments to be discussed
below we have also performed the numerical analysis including several
refinements that pertain to our specific experimental situation. These effects
are: (a) a difference $\left(  \delta y\right)  $ in gravitational sag between
the IP and the TOP trap; (b) an exponential switching transient of the current
in the TOP coils and correspondingly in the $\mathbf{B}_{m}$ field $\left(
\tau_{1/e}=3~\mu\mathrm{s}\right)  $; (c) a switching transient of
$\sim0.5~\mathrm{ms}$ in the offset field from $B_{0}=9.5\times10^{-5}%
~\mathrm{T}$ at the $t=0$ to the final value $B_{0}=3.1\times10^{-5}%
~\mathrm{T}$.

The initial gravitational sag in the IP trap is $1.2\mathrm{~\mu m.}$ When
switching on the TOP, the sag $\Delta y$ jumps in $\sim3~\mu\mathrm{s}$ to
$1.7~\mathrm{\mu m}$ and settles in $\sim0.5~\mathrm{ms}$ to its final value
$1.6~\mathrm{\mu m}$ due to the decrease of $B_{0}$. Thus the gravitational
sag increases jump wise and settles at $\delta y=0.4~\mathrm{\mu m}$. During
the same transient the radius of the stationary micromotion grows from
$\rho_{0}=0.21~\mathrm{\mu m}$ to $\rho_{0}=0.33~\mathrm{\mu m}$ and $\Omega$
increases by about $5\%$.

The dotted red traces in Fig.\thinspace\ref{theortraject} correspond to the
numerical calculation including all the above refinements relevant to the
experiments. We have also investigated the effects of gravity, $\mathbf{B}%
_{m}$-switching and $B_{0}$-switching separately. We find that the main effect
of the settling time of $B_{0}$ is to reduce the amplitude along the major
axis by $\sim35\%$. The combined effect of changing gravitational sag and
$\mathbf{B}_{m}$ transient is to slightly increase the $x$ amplitude as well
as to produce a slight tilt angle of the trajectory (see right-most panel of
Fig.\thinspace\ref{theortraject}).

The tilt angle $\theta$ of the macromotion also follows from a fit of
Eq.\thinspace(\ref{r_approx}) to the numerical results: for known values of
$X_{0}$, $Y_{0}$, $\varphi_{x}$ and $\varphi_{y}$ the angle of rotation
$\vartheta$ to align the coordinate system along the major and minor axis is
given by%
\begin{equation}
\vartheta=\tfrac{1}{2}\tan^{-1}\left[  2\sin(\varphi_{x}-\varphi_{y}%
)X_{0}Y_{0}/(Y_{0}^{2}-X_{0}^{2})\right]  \label{eq:rotation}%
\end{equation}
For $\phi_{m}=0$ the tilt angle equals the rotation angle $(\theta=\vartheta)$.

The results of a fit of Eq.\thinspace(\ref{r_approx}) to the numerical results
including only the refinements (a) and (b), as well as a fit including all
three refinements (a), (b) and (c) are also given in
Table~\ref{table:FitResultsA}. Extending the analytical model to include the
refinements (a) and (b) is straightforward and given in detail in Appendix
\ref{app:Delayed Sudden-Step Model}. The expressions for the amplitudes and
phases depend on the model parameter $\tau_{0}$ and are given by
Eqs.\thinspace(\ref{X0})-(\ref{Phi-y}) of the appendix. The model parameter
$\tau_{0}$ is chosen by ensuring that the value of the tilt angle $\theta$ of
the model reproduces that of a fit to the numerical solution for zero settling
time, $\theta=0.024\pi$. This results in $\tau_{0}=3.5~\mu\mathrm{s}$.
Excellent agreement is obtained with the numerical model as is shown in
Table~\ref{table:FitResultsA}. Insight in the cause of the reduction of the
major-axis amplitude associated with the settling behavior of $B_{0}$ can also
be gained using the analytical model. As discussed in Appendix
\ref{app:Delayed Sudden-Step Model} the major refinement is change the launch
speed corresponding to the initially smaller value of $\rho_{0}$. Although
this refinement captures the origin of the $35\%$ reduction of the major axis
amplitude, Table~\ref{table:FitResultsA} shows that the overall agreement with
the numerical model is less favorable.

\subsection{Phase jumps}

\label{sec:Phase jumps} Let us now analyze how the macromotion can be
quenched. For a one-dimensional, rapidly-oscillating potential it was
demonstrated in Ref.\thinspace\cite{Rid1} that the amplitude of the
macromotion can be quenched by an appropriate phase-jump of the modulation
field. For the 2D motion in a TOP, the success of such an approach is not
\emph{a priori} obvious because the phase jumps for the $x$- and $y$ motion
cannot be selected independently. Yet, as will be shown below, also for the
TOP it is possible to quench both the $X_{0}$- and $Y_{0}$ amplitudes more or
less completely by imposing a single phase jump $\Delta\phi_{m}$ to the
$\mathbf{B}_{m}$ field.%
\begin{figure}[ptb]%
\centering
\includegraphics[
height=4.848cm,
width=8.2642cm
]%
{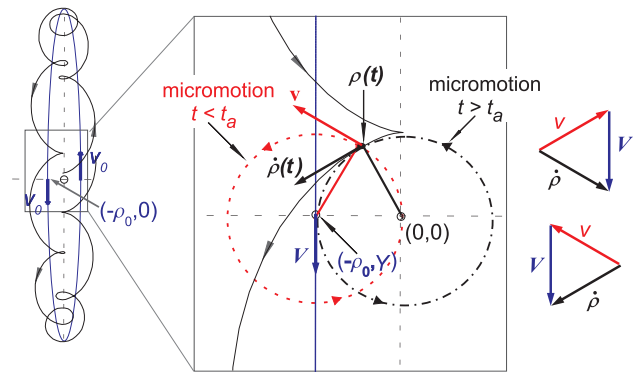}%
\caption{(color online) Explanatory diagram for the phase jump. Left: cloud
trajectory (black solid line) along with macromotion trajectory (blue dotted
line) The black dashed lines are the symmetry axes of the trap and the blue
arrows show the macromotion velocity on crossing the $x$-axis. Middle:
Expanded view of boxed region of the left panel; $\boldsymbol{\rho}\left(
t\right)  $ is the position of the cloud at the time of the phase jump. The
red dashed (black dot-dashed) circle is micromotion just before (after) the
phase jump at $t=t_{a}$. Right: micromotion $(\mathbf{v})$ and
macromotion$(\mathbf{V})$ velocity vectors add up to the total velocity vector
$\boldsymbol{\dot{\rho}}\left(  t\right)  $. }%
\label{fig:jump}%
\end{figure}

For clarity we first restrict ourselves to the case $\phi_{m}=0$ and neglect
the effects of gravity and switching transients. This means that the cloud is
launched at $t=0$ in the vertical $y$ direction with a speed that is equal to
$v_{0}$, the micromotion speed. As can be seen from the trajectory depicted at
the left of Fig.\thinspace3 the macromotion speed will again be equal to
$v_{0}$ when the cloud returns close to the origin after an integer number of
macromotion half-periods. The total velocity $\boldsymbol{\dot{\rho}}\left(
t\right)  $ is the vector sum of the micro- and macromotion velocities and
this quantity varies rapidly on a time scale of the micro-motion period.

The essence of the quenching procedure is to apply the phase jump at a time
$t_{a}$ chosen in the interval $t_{n}-\Delta t<t<t_{n}+\Delta t$ around times
$t_{n}=n\left(  \pi/\Omega\right)  $ corresponding to a multiple of the
macromotion half-period. We choose $t_{a}$ such that $\boldsymbol{\dot{\rho}%
}\left(  t_{a}\right)  $ has a magnitude equal to $v_{0}$. When the cloud
returns at the $x$ axis the micro- and macromotion speeds are both $v_{0}$ and
hence the resultant total velocity can only be equal to $v_{0}$ if the angle
between the macro- and micromotion directions is either $2\pi/3$ or $-2\pi/3$
corresponding to two distinct micromotion phases $\phi_{a}\equiv\phi
(t_{a})=\omega t_{2n-1}+\phi=\pm\pi/3$ (see Fig.\thinspace3-right). In other
words the micro- and macromotion velocity vectors form an equilateral
triangle. For each of these cases a corresponding phase jump exists,
$\Delta\phi_{m}=\pm\pi/3$ respectively, such that $\boldsymbol{\hat{\rho}}%
_{m}$ is set perpendicular to $\boldsymbol{\dot{\rho}}\left(  t_{a}\right)  $,
which sets the macromotion velocity to zero. The result is pure micromotion if
the orbit into which the particle is kicked is centered around the origin. For
each of the two choices of $\phi_{a}$, pure micromotion results only if the
macromotion position at the time of the phase jump is equal to $(\pm\rho
_{0},0)$, where the $+$ $(-)$ sign applies for even (odd) $n$. Complete
quenching can be achieved only for specific choices of the ratio
$\omega/\Omega$. The change of orbit upon a phase jump is explained
pictorially in the middle of Fig.\thinspace3.

We now generalize to the case where the ratio $\omega/\Omega$ is not precisely
fine tuned and allow for the possibility that the macromotion speed deviates
slightly from the value $v_{0}$ assumed above. One can show that, also in this
case, the maximal reduction in marcromotion energy resulting from a phase jump
is achieved when the jump is applied at a time $t_{a}$ when $\boldsymbol{\dot
{\rho}}\left(  t_{a}\right)  $ has a magnitude equal to $v_{0}$. The value of
$\Delta\phi_{m}$ is again selected such as to set $\boldsymbol{\hat{\rho}}%
_{m}$ perpendicular to $\boldsymbol{\dot{\rho}}\left(  t_{a}\right)  $. By a
reasoning similar to the case described above we find that the condition of an
equilateral triangle of the three velocity vectors is now replaced by one that
is isosceles-triangle condition with the micro-motion velocity and
$\boldsymbol{\dot{\rho}}\left(  t_{a}\right)  $ both having a magnitude
$v_{0}$. This in turn means that the magnitude of the phase jump will deviate
slightly from the values $\pm\pi/3$ found above. Also, the nearest distance to
the $x$ axis at which the isosceles-triangle condition can be met is in
general not equal to zero. This means that some residual macromotion will be
present after the phase jump, with an amplitude given by the distance to the
origin of the center of the circular orbit into which the cloud is transferred
by the phase jump. One can show that there is always a choice possible where
the isosceles-triangle condition is satisfied such that this distance is
approximately $2\rho_{0}$ or less. As a consequence, even in the worst case,
the macromotion amplitude is reduced from $(\omega/\Omega)\rho_{0}$ to an
amplitude of order $\rho_{0}$.

The criterion that the acceleration be set perpendicular to the total velocity
at the time that the macromotion speed is equal to $v_{0}$ can be expressed by
the following equation:%

\begin{equation}
\Delta\phi_{m}=\arctan\left[  \frac{\dot{\rho}_{y}\left(  t_{a}\right)  }%
{\dot{\rho}_{x}(t_{a})}\right]  -\phi\left(  t_{a}\right)  +(-1)^{k}\frac{\pi
}{2} \label{eq:phasejump}%
\end{equation}
where $\dot{\rho}_{x}\left(  t_{a}\right)  =-v_{0}\sin\phi\left(
t_{a}\right)  -V_{0,x}\sin(\Omega t_{a}+\varphi_{x})$ and $\dot{\rho}%
_{y}\left(  t_{a}\right)  =v_{0}\cos\phi\left(  t_{a}\right)  +V_{0,y}%
\cos(\Omega t_{a}+\varphi_{y})$ are $x$- and $y$ components of
$\boldsymbol{\dot{\rho}}$ at time $t_{a}$ and $k=1$ for $\dot{\rho}_{x}%
(t_{a})$ $>0$ and $k=0$ for $\dot{\rho}_{x}(t_{a})<0$. We return to selection
of the jump time and the use of Eq.\thinspace(\ref{eq:phasejump}) when
discussing the measurement procedure in Section
\ref{sec:Measurement procedure}.

\begin{figure}[ptb]%
\centering
\includegraphics[
height=8.4455cm,
width=8.2621cm
]%
{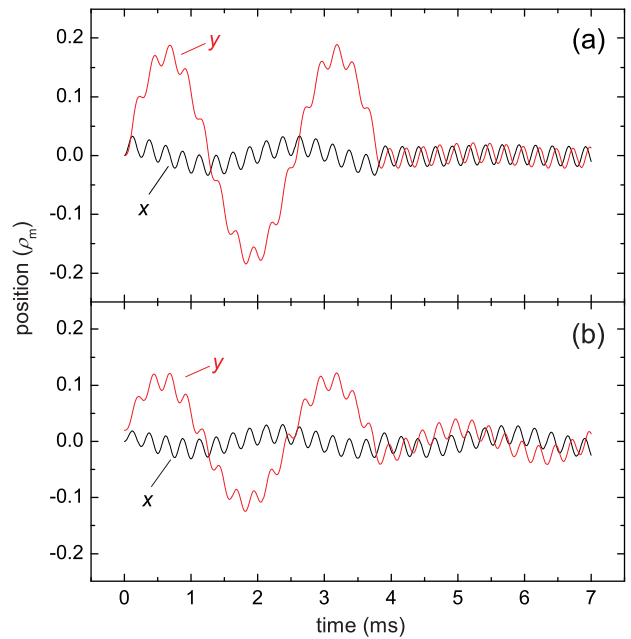}%
\caption{(color online) Numerically calculated radial trajectories in the $x$-
(black) and $y$ (red) direction for the same trap parameters as used for
Fig.~\ref{theortraject}, with a quenching phase jump $\Delta\phi_{m}$ applied
at optimized $t=t_{a}$ $\simeq$ $t_{3}$ (three macromotion half-periods). (a)
instant switching, no gravity: $\Delta\phi_{m}=-\pi/3$, $t_{a}$
$=3.834\,\mathrm{ms}$; (b) including switching transients and gravity:
$\Delta\phi_{m}=-0.22\pi$, $t_{a}$ $=$ $3.834\,\mathrm{ms}$.}%
\label{x-y-motiontheory}%
\end{figure}

Examples of the numerical calculations of the quenching procedure are shown in
Fig.\thinspace\ref{x-y-motiontheory}.%
The near complete quenching of the macromotion shown in panel (a) is obtained
for $\delta y=0$ and $\tau=0$ with phase jump $\Delta\phi_{m}=-\pi/3$ at time
$t_{a}=3.834~\mathrm{ms}$ in the time interval around $t_{3}=3\pi/\Omega$. In
Fig.\thinspace\ref{x-y-motiontheory}b the refinements (a), (b) and (c) are
included in the simulation of the experiment. In this case the phase jump had
to be adjusted to $\Delta\phi_{m}=-0.22\pi$ for maximum quenching. Note that
the quenching is less complete. By adjusting, at constant $\Omega$, the
micromotion frequency to $\omega=4.068~\mathrm{kHz}$ and the jump time to
$t_{a}=3.769~\mathrm{ms}$, complete quenching similar to that shown in panel
(a) was obtained also when including all refinements in the numerical model.

\section{Experimental\label{experimentaldetails}}

\subsection{Apparatus}

The experiments are done with$\,$the apparatus described in detail in
\cite{DiecTh} and \cite{BuggleTh}. We produce a \ BEC of $2.5\times10^{5}$
atoms of $^{87}$Rb in the $|F=2,m_{F}=2\rangle$ state in a Ioffe-Pritchard
trap using radio-frequency (rf) evaporative cooling. The symmetry axis ($z$
axis) of the trap lies horizontal with trap frequencies ($\Omega_{\rho}%
/2\pi=455(5)~$\textrm{Hz}, $\Omega_{z}/2\pi=21~$\textrm{Hz}) and the magnetic
field offset $B_{0}=9.5(3)\times10^{-5}~\mathrm{T}$, $\alpha=3.53~\mathrm{T/m}%
$ and $\beta=266~\mathrm{T/m}^{2}$. The Thomas-Fermi radius of the BEC is
$2.2~\mu\mathrm{m}$. The TOP field is produced by two pairs of coils, one in
the $x$ direction, the other in the $y$ direction as described previously in
\cite{Tiecke03}. The coils consist of only two windings to keep the inductance
low. The current for the TOP is generated by a TTI 4 channel arbitrary
waveform generator (TGH 1244), amplified by a standard audio-amplifier (Yamaha
AX-496). The current used is $I_{m}=3.0~\mathrm{A}$ and the field produced is
$B_{m}=6.8(2)\times10^{-5}~\mathrm{T}$. All measurements in the TOP are done
with $\Omega/2\pi=394(4)~\mathrm{Hz}$ ($B_{0}=3.1\times10^{-5}~\mathrm{T}$).
Detection is done by time-of-flight absorption imaging along the $z$ axis
using a one-to-one transfer telescope to image the $xy$ plane onto a Princeton
TE/CCD-512EFT CCD camera with $15~\mu\mathrm{m}$ pixel resolution. All
measurements are carried out with the same flight time $\Delta t_{\mathrm{TOF}%
}=23~\mathrm{ms}$, giving rise to an expanded cloud radius of $\sim
140~\mu\mathrm{m}$.

\subsection{Measurement procedure\label{sec:Measurement procedure}}

Our experiments on phase-jump-controlled motion in a TOP trap are done with
the $\mathbf{B}_{m}$ field operated at $\omega/2\pi=4\mathrm{~kHz}$. This
frequency is sufficiently high $\left(  \omega/\Omega\gtrsim10\right)  $ to
satisfy the `TOP condition' Eq.\thinspace(\ref{TOPcondition}). The frequency
is chosen lower than in a typical TOP to ensure that the speed of the
stationary micromotion, $9~\mathrm{mm/s}$ as estimated with Eq.\thinspace
(\ref{vel-0}), is accurately measurable. In the experiments we start with an
equilibrium BEC in the IP trap described above. At $t=0$ we switch on the
$\mathbf{B}_{m}$ field, using $B_{0}$ to tune the measured trap frequency to
$\Omega/2\pi=394~\mathrm{Hz}$. As the trap minimum shifts down by $\delta
y=0.40~\mu\mathrm{m}$, the initial position of the cloud is slightly above the
trap center. The $1/e$-switching time of the $\mathbf{B}_{m}$ field was
measured to be $\tau\thickapprox3~\mu\mathrm{s}$, which corresponds to
$\omega\tau\thickapprox0.08$. When changed, the $B_{0}$ field settles to a new
value after a damped oscillation with a frequency of 650 Hz and a damping time
$\tau^{\prime}$ of 0.56 ms. This corresponds to $\Omega\tau^{\prime
}\thickapprox0.2$. The velocity $\boldsymbol{\dot{\rho}}$ of the BEC in the
radial plane at the time of release is determined by time-of-flight absorption
imaging along the $z$ axis. For the chosen flight time of $23~\mathrm{ms,}$ a
speed of $1~\mathrm{mm/s}$ corresponds to a displacement of $23~\mu\mathrm{m}$
with respect to a cloud released from the same position at zero velocity. A
cloud released at rest at time $t_{rel}$ is imaged at position \ $\mathbf{R}%
_{0}=\boldsymbol{\rho}(t_{rel})+\frac{1}{2}\boldsymbol{\ddot{\rho}}%
_{g}\boldsymbol{~}\Delta t_{\mathrm{TOF}}^{2}$, where $\boldsymbol{\ddot{\rho
}}_{g}$ is the gravitational acceleration. For a finite release velocity
$\boldsymbol{\dot{\rho}}(t_{rel})$ the cloud will be imaged at $\mathbf{R}%
=\boldsymbol{\dot{\rho}}(t_{rel})\Delta t_{\mathrm{TOF}}+\mathbf{R}_{0}$.

In practice we may neglect the small variation in the release position due to
the macromotion, approximating $\boldsymbol{\rho}(t_{rel})\simeq\rho(0)$,
because this variation is smaller than the shot-to-shot reproducibility of the
cloud position. From the model analysis of Section
\ref{sec:micromotion and macromotion} the variation in release position due to
the macromotion is estimated to be $\delta\boldsymbol{\rho}(t_{rel}%
)\lesssim(\omega/\Omega)\rho_{0}\approx4~\mu\mathrm{m}$. The centroid of the
image of the expanded cloud is determined using a simple Gaussian fitting
procedure and has a shot-to-shot reproducibility of $\sim8\mathrm{~\mu m}$,
small as compared to the $140~\mu\mathrm{m}$ radius of the expanded cloud. No
improvement in shot-to-shot reproducibility was found by changing to a higher
magnification. Since our measurements depend only on the position of the cloud
center they are insensitive to fluctuations in atom number or density.%
\begin{figure}[ptb]%
\centering
\includegraphics[
height=5.972cm,
width=8.2621cm
]%
{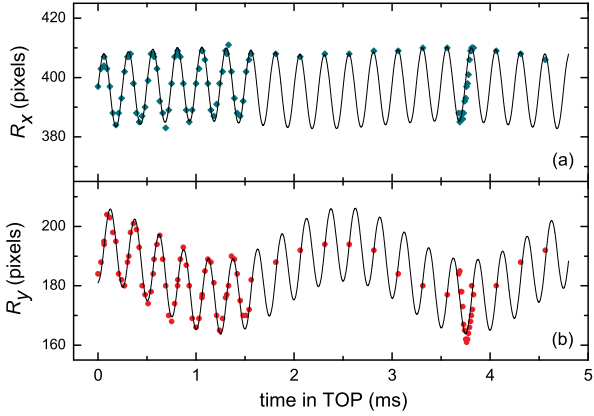}%
\caption{(color online) The centroid position after $23~\mathrm{ms}$ TOF
plotted in camera pixel units against holding time in the TOP trap: upper
datatset: $R_{x}$; lower dataset: $R_{y}$. The solid lines represent the fit
of Eqs.\thinspace(\ref{velFit-x}) and (\ref{velFit-y}) to the data. Note that
by a stroboscopic measurement at $0.25~\mathrm{ms}$ intervals the micromotion
is eliminated. Each point represents a single measurement.}%
\label{rawdataandfits}%
\end{figure}
To reconstruct the motion of the condensate in the trap we image the cloud at
$t=t_{i}$, where $t_{i}$ is the holding time in the TOP. We obtain the release
velocity by measuring the $x$ and $y$ components of the cloud centroid
$(R_{x},R_{y})$. A typical set of data is shown in Fig.\thinspace
\ref{rawdataandfits}. The micromotion is recognized as the rapid modulation on
the slow macromotion. As the frequency of the micromotion is accurately known
we avoid aliasing by sampling the motion in steps of $0.025~\mathrm{ms}$, much
shorter than the micromotion period. If we wish to look only at the
macromotion in a stroboscopic manner, we can sample precisely at the
micromotion period of $0.25~\mathrm{ms}$, with best results obtained when
sampling on the crests of the micromotion. Fitting the expressions%
\begin{align}
R_{x}  &  =-v_{0}\Delta t_{\mathrm{TOF}}\sin(\omega t+\phi)-\nonumber\\
&  \ \ \ \ \ \ \ \ \ \ \ \ \ -V_{0,x}\Delta t_{\mathrm{TOF}}\sin(\Omega
t+\varphi_{x})+R_{0,x}\label{velFit-x}\\
R_{y}  &  =v_{0}\Delta t_{\mathrm{TOF}}\cos(\omega t+\phi)+\nonumber\\
&  \ \ \ \ \ \ \ \ \ \ \ \ \ +V_{0,y}\Delta t_{\mathrm{TOF}}\cos(\Omega
t+\varphi_{y})+R_{0,y} \label{velFit-y}%
\end{align}
to the data, using the TOP frequency $\omega$ and $\Delta t_{\mathrm{TOF}}$ as
known parameters, we obtain the amplitudes $v_{0}$, $V_{0,x}$, $V_{0,y}$ as
well as the macromotion frequency $\Omega$ and the phases $\phi,\varphi
_{x},\varphi_{y}.$ Note that the fit also yields the reference position
$\mathbf{R}_{0}=\{R_{0,x},R_{0,y}\}$ but this information is superfluous for
the reconstruction of the in-trap motion. Once these quantities are determined
the motion of the condensate in the TOP trap is readily reconstructed with
Eq.\thinspace(\ref{r_approx}).%
\begin{figure}[ptb]%
\centering
\includegraphics[
trim=0.217032in 0.216686in 0.217448in 0.363322in,
height=5.6451cm,
width=8.2621cm
]%
{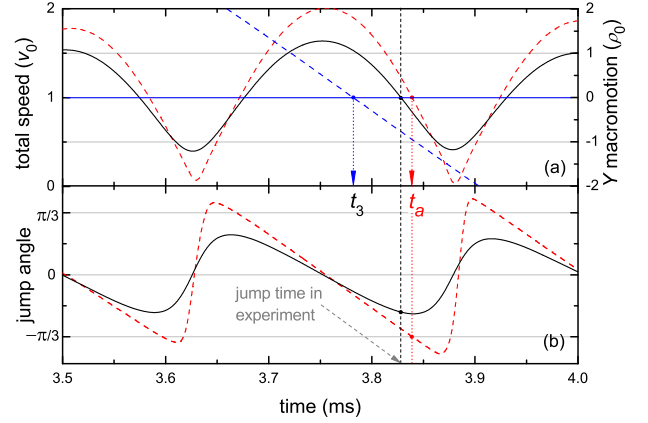}%
\caption{(color online) Illustration of how to choose the optimal phase jump
and its timing. In both panels: solid black curve - experimental conditions;
dashed red curve - analytical model of Section \ref{sec:Phase jumps} for the
case of instant switching - no gravity. (a) The total speed of the cloud in
units of the micromotion speed $v_{0}$ (optimal phase jump time $t_{a}$
corresponds to $\dot{\rho}(t_{a})=v_{0}$); the dashed blue line (scale on
right) shows the $Y$ component of the macromotion position crossing zero at
$t=t_{3}$ (stationary micromotion can be achieved by adjusting $\omega$ such
that $t_{3}=t_{a}$); (b) The optimal phase jump as a function of jump time as
calculated by Eq.\thinspace(\ref{eq:phasejump}).}%
\label{timing}%
\end{figure}

To investigate the effect of phase jumps, we implement the approach described
in Section \ref{sec:Phase jumps}. First we determine for given $\omega$ and
$\Delta t_{\mathrm{TOF}}$ all parameters to reconstruct the motion with the
method just described. This enables us to determine the time intervals
$t_{n}-\Delta t<t<t_{n}+\Delta t$, where the cloud returns close to the
origin, and choose within this interval the time $t_{a}$, where the total
velocity $\boldsymbol{\dot{\rho}}\left(  t_{a}\right)  $ has magnitude $v_{0}$
as shown in Fig.\thinspace\ref{timing}a. The red dashed lines correspond to
the analytical model of Section \ref{sec:Phase jumps} for the case of instant
switching - no gravity (the case of Fig.\thinspace\ref{x-y-motiontheory}a).
The black solid lines correspond to the calculation including all relevant
experimental constraints. The phase jump $\Delta\phi_{m}$ that sets
$\boldsymbol{\hat{\rho}}_{m}$ perpendicular to $\boldsymbol{\dot{\rho}}\left(
t_{a}\right)  $ is given by Eq.\thinspace(\ref{eq:phasejump}). This optimal
phase jump $\Delta\phi_{m}$ is plotted versus $t_{a}$ in a time interval
around $t_{3}=3\pi/\Omega$ in Fig.\thinspace\ref{timing}b. For the case of
instant switching - no gravity the optimum phase jump is seen to be
$\Delta\phi_{m}=-\pi/3$. At the chosen time $t_{a}$ we vary the phase jump
$\Delta\phi_{m}$ about the value suggested by Eq.\thinspace(\ref{eq:phasejump}%
) in search for optimal quenching. To reconstruct the residual macromotion, we
hold the cloud for a variable additional time $t_{b}$, before TOF imaging at
time $t=t_{a}+t_{b}$.

\section{Results and discussion}

\label{results}

\begin{figure}[ptb]%
\centering
\includegraphics[
trim=0.214174in 0.257684in 0.236814in 0.204854in,
height=6.3284cm,
width=8.2642cm
]%
{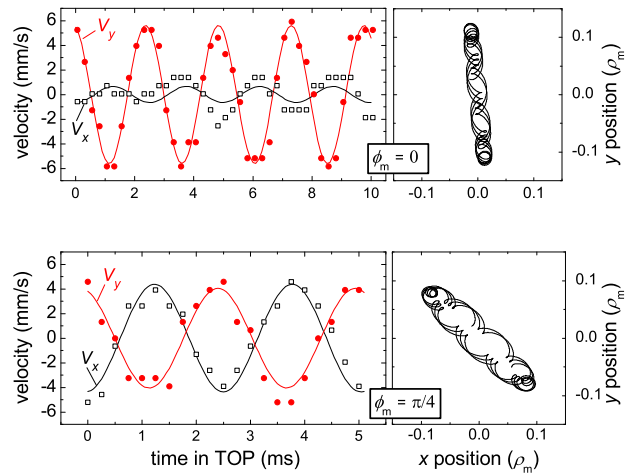}%
\caption{(color online) The panels on the left show the macromotion velocities
(taken with the stroboscopic method) of the cloud centroid $x$- (black) and
$y$- (red) versus time for $\phi_{m}=0$ and $\pi/4$. The solid curves are fits
of Eqs.\thinspace(\ref{velFit-x}) and (\ref{velFit-y}) to the data. The panels
on the right represent the reconstructed trajectories in parametric form (in
units of the TOP radius $\rho_{m}=19.5$ $\mu\mathrm{m}$. The difference in
aspect ratio is caused by the gravity shift.}%
\label{figure-parametric}%
\end{figure}

In this section we show the results obtained with the experimental procedure
described in the previous section. We measured the macromotion induced by
switching on the $\mathbf{B}_{m}$ field for three values of the initial TOP
phase, $\phi_{m}=0,\pi/4,\pi/2$. For $\phi_{m}=0$ part of the raw data are
shown in Fig.\thinspace\ref{rawdataandfits}. In Fig.\thinspace
\ref{figure-parametric} we show the measured velocity of the macromotion
obtained with the stroboscopic method. The upper and lower panels correspond
to $\phi_{m}=0$ and $\pi/4$ respectively. The data for $\phi_{m}=\pi/2$ are
not shown but are similar to those for $\phi=0$ but with the roles of $x$ and
$y$ interchanged.

The solid lines in the left panels of Fig.\thinspace\ref{figure-parametric}
are obtained by fitting Eqs.\thinspace(\ref{velFit-x}) and (\ref{velFit-y}) to
the full data including micromotion and provide the input for calculating the
amplitudes. Using the known TOP frequency $\omega/2\pi=4~\mathrm{kHz}$ and
flight time $\Delta t_{\mathrm{TOF}}=23~\mathrm{ms}$, the fit yields for the
velocity amplitudes, phases, and frequency: $v_{0}=7.6(2)~\mathrm{mm/s,}$
$V_{0,x\text{ }}=0.7(2)~\mathrm{mm/s}$, $V_{0,y}=5.6(2)~\mathrm{mm/s,}$
$\phi=1.00(1)\pi$, $\varphi_{x}=0.5(2)\pi$, $\varphi_{y}=$ $0.05(2)\pi$, and
$\Omega/2\pi=394(4)~\mathrm{Hz}$. The corresponding in-trap amplitudes are
$\rho_{0}\equiv v_{0}/\omega=0.30(1)~\mu\mathrm{m,}$ $X_{0}\equiv V_{0,x\text{
}}/\Omega=0.28(7)~\mu\mathrm{m}$, and\ $Y_{0}\equiv V_{0,y\text{ }}%
/\Omega=2.3(1)~\mu\mathrm{m}$.%

The right panels in Fig.\thinspace\ref{figure-parametric} are parametric plots
of the trajectories obtained by reconstructing the motion in the trap from the
velocity fits described above. The trajectories provide a useful way to see
the effect of the initial phase of the applied $\mathbf{B}_{m}$ field and in
addition the upper panel can be directly compared with the theoretical
prediction shown in Fig.\thinspace\ref{theortraject}. As expected, the
orientation of the major axis of the macromotion is dependent on the initial
phase $\phi_{m}$ of the $\mathbf{B}_{m}$ field. The small tilt $\theta$ away
from the direction perpendicular to $\boldsymbol{\hat{\rho}}_{m}$ is clearly
visible and consistent with the calculations for a finite switch-on time and
the presence of gravity. The value obtained for $\rho_{0}$ is slightly smaller
than the value calculated with Eq.\thinspace(\ref{rho-0}) but in view of
experimental uncertainties certainly consistent with the value of $\alpha$.
The results for $\varphi_{x}/\pi$, $\varphi_{y}/\pi$, $X_{0}/\rho_{0}$%
,$Y_{0}/\rho_{0}$ and the tilt angle $\theta$ obtained for $\phi_{m}=0$ and
$\pi/4$ are given in Table~\ref{table:FitResults}. For comparison, also the
numerical results are included.%

\begin{table}[b] \centering
\caption{Experimental results (exp) for macromotion induced by the switch-on of the TOP field for $\phi_{m}=0,\pi/4$. The data are compared with the results of the numerical calculation of Section \ref{sec:Numerical analysis} (num). In all cases the tilt angle has be calculated with the aid of Eq.\,\ref{eq:rotation}}%
\begin{tabular}
[c]{c|c|c|c|c|c|c}\hline\hline
& $\phi_{m}$ & $\theta/\pi$ & $\varphi_{x}/\pi$ & $\varphi_{y}/\pi$ &
$X_{0}/\rho_{0}$ & $Y_{0}/\rho_{0}$\\\hline
{\small exp} & $0$ & $0.04(2)$ & $0.5(2)$ & $0.05(2)$ & $0.9(2)$ &
$7.7(3)$\\\hline
{\small num} & $0$ & $0.016$ & $0.20$ & $0.06$ & $0.82$ & $6.5$\\\hline
{\small exp} & $\pi/4$ & $0.04(2)$ & $0.49(2)$ & $0.12(2)$ & $6.6(4)$ &
$5.2(3)$\\\hline
{\small num} & $\pi/4$ & $0.013$ & $0.47$ & $0.04$ & $4.95$ & $4.55$%
\end{tabular}
\label{table:FitResults}%
\vspace{3mm}
\end{table}%

\begin{figure}[ptb]%
\centering
\includegraphics[
trim=0.212714in 0.199177in 0.199334in 0.199631in,
height=4.0108cm,
width=8.2621cm
]%
{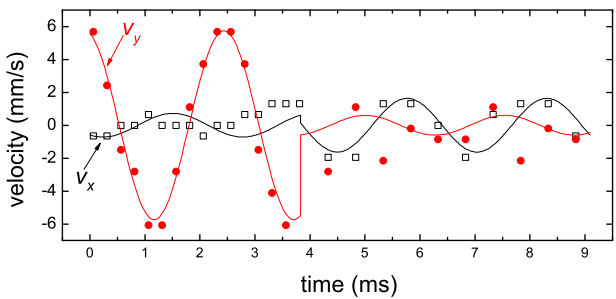}%
\caption{(color online) Measured and calculated velocity of the macromotion
before and after a phase jump of $\Delta\phi_{m}=-0.22\pi$ at $t_{a}%
=3.83~\mathrm{ms}$ for initial phase $\phi_{m}=0$ . The open black squares
(solid red circles) correspond to the measured $V_{x}$ $\left(  V_{y}\right)
$ velocity component (for $t_{a}<3.83~\mathrm{ms}$ the data coincide with
those of Fig.\thinspace\ref{figure-parametric}). Each point represents a
single measurement. The solid lines correspond to the numerical model without
any adjustable parameter as described in the text. }%
\label{quenching}%
\end{figure}

We now turn to the results of a quenching experiment. The time $t_{a}%
\simeq3.83~\mathrm{ms}$ and magnitude $\Delta\phi_{m}=-0.22\pi$ of the phase
jump have been chosen to meet the conditions necessary to quench the
macromotion as introduced in Section \ref{sec:Measurement procedure} and
illustrated in Fig.\thinspace\ref{timing}. In Fig.\thinspace\ref{quenching} we
show velocity data taken with the stroboscopic method. For $t<3.83~\mathrm{ms}%
$ the data coincide with those shown in the upper panel of Fig.\thinspace
\ref{figure-parametric} but the solid lines are not a fit but represent the
macromotion velocity predicted by the numerical calculation on the basis of
the experimental parameters. These velocity curves correspond to the
macromotion part of the position plot Fig. \ref{x-y-motiontheory}b and have no
adjustable parameters. Both experiment and theory show pronounced reduction in
the amplitude of the macromotion. Although, the phases of the quenched motion
cannot be determined convincingly with our signal to noise ratio, the
agreement between theory and experiment is satisfactory.%

In general a jump in the micromotion phase produces an abrupt change in
macromotion phase and amplitude. For the case illustrated in Fig.
\ref{quenching} we obtain a reduction of more than a factor of $5$ in the
amplitude of oscillation in the $y$ direction at the expense of only a slight
increase of the amplitude in the $x$ direction. As a result the macromotion is
reduced to the size of the micromotion. The energy associated with the
macromotion is consequently reduced by a factor of about 15, reducing it to a
small fraction of the micromotion energy. This demonstrates that the initial
sloshing motion of the cloud can be efficiently quenched by applying an
appropriate phase jump angle. As pointed out in the last paragraph of Section
\ref{sec:Phase jumps} we expect that it should be possible to suppress the
macromotion almost completely by adjusting the micromotion frequency such that
$t_{3}=t_{a}$.%
\begin{figure}[ptb]%
\centering
\includegraphics[
trim=0.216679in 0.199046in 0.199459in 0.199294in,
height=5.4385cm,
width=8.2621cm
]%
{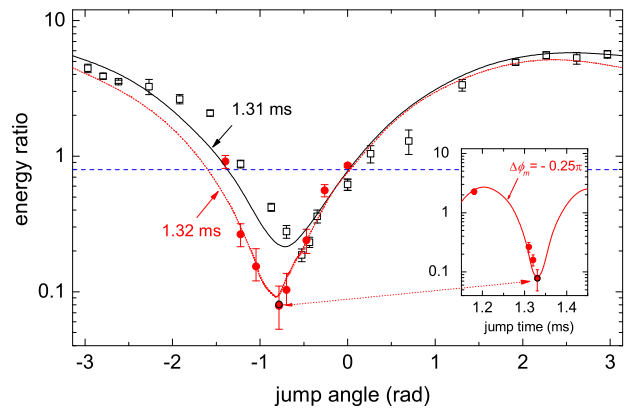}%
\caption{(color online) Ratio of macromotion energy over micromotion energy
following a phase jump plotted against $\Delta\phi_{m}$ at $1.32$~\textrm{ms}
(open black squares) and $1.33$~\textrm{ms} (red circles). Each data point is
obtained from fits as described in the text. The horizontal blue dashed line
shows the initial value of the energy before the phase jump. The solid black
lines and dotted red are the numerical calculation for $1.31$~\textrm{ms} and
$1.32$~\textrm{ms} respectively. The inset shows the dependence on jump time
for fixed value of $\Delta\phi_{m}$.}%
\label{vary phase jump}%
\end{figure}

Even a small variation in the phase jump magnitude or its timing can result in
a substantial difference in quenching efficiency. This is illustrated in
Fig.\thinspace\ref{vary phase jump}, where we plot the ratio of macro- and
micromotion energy,
\begin{equation}
\frac{E_{macro}}{E_{micro}}=\frac{V_{0,x\text{ }}^{2}+V_{0,y}^{2}}{v_{0}^{2}},
\end{equation}
as the phase jump is varied in steps of $10$ degrees, for $t_{a}%
=1.32~\mathrm{ms}$ and $1.33~\mathrm{ms}$, where the position and velocity
criteria are well satisfied. For most phase jumps $\Delta\phi_{m}$ the result
is an \textit{increase} in energy. The drawn lines are the predictions from
the numerical model for the same conditions at $t_{a}=1.31~\mathrm{ms}$ and
$t_{a}=1.32~\mathrm{ms}$. The plot for $t_{a}=1.32~\mathrm{ms}$ shows a deeper
reduction than that for $t_{a}=1.31~\mathrm{ms}$, as well as a shifted optimal
$\Delta\phi_{m}$. The common shift of $\sim0.01~\mathrm{ms}$ between the data
and the numerical results remains unexplained.

\section{Summary and conclusion}

\label{Discussion and Outlook}

We have shown that a cold atomic cloud initially at rest at the minimum of the
effective potential of a TOP trap, acquires a macroscopic sloshing motion, in
addition to near circular micromotion, when the TOP is suddenly switched on.
The energy associated with this macromotion is of the same order as the energy
of the micromotion and the amplitude of the former is larger than that of the
latter by a factor $\sim\omega/\Omega$. We have theoretically described the
phenomenon and the predictions compare well with our experimental results.

As the micromotion is shared in common mode by all trapped atoms, the
associated energy does not affect the thermodynamics of the cloud in any way.
In contrast, the macromotion energy is generally unwanted and potentially
harmful. Fortunately, as we have shown, it is possible to quench this
macromotion almost completely and instantly, by applying a suitable and
properly timed phase jump to the rotating magnetic field that defines the TOP.
We have shown theoretically that this procedure works, even for the 2D case of
the TOP, which is an extension of previous theory describing similar phenomena
in 1D \cite{Rid1, Rid2}. We have presented a framework which allows a
deterministic procedure for choosing the optimal parameters for the phase
jump. Our experiments corroborate the theoretical model for the TOP in a
quantitative manner.

The macromotion induced by the switch-on and the subsequent possibility to
alter this motion by phase jumps have several consequences, some of which we
now briefly mention. For example, the sloshing motion may affect the time of
flight imaging once the fields have been switched off. When comparing
TOF-images for different holding times it is in general not sufficient to
synchronize the release time to the micromotion period. The position after TOF
can be easily polluted by the non-zero macromotion, which evolves
asynchronously with the micromotion. The time scales in this experiment are on
the order of a few macromotion periods. The physics of interest of the cloud
is usually seen on much longer time scales of hundreds of such periods. On
these longer time scales, the presence of even small anharmonicities can lead
to the conversion of macromotion energy into heat. The macromotion can be of
order of the chemical potential which can have consequences for the stability
of the condensate.

The possibility to excite or quench macromotion by phase jumps of the rotating
field is a valuable feature of the TOP trap that has received little attention
in the literature. Our work shows that this feature is well understood and can
be applied in a well-controlled manner. We have primarily focussed on
quenching with a single phase jump. However, the reverse effect in which the
macromotion is excited may prove equally useful in some experiments. Also the
consequences for multiple phase-jump applications deserve attention in this
respect. We established numerically that it should be possible to excite or
deexcite large macromotion with a series of $\pi$ phase jumps at intervals of
the macromotion half-period. At each of these phase jumps, either component of
the macromotion velocity can be increased or decreased by $\sim2v_{0}$. Being
outside the primary focus of this paper, we do not further elaborate on this
interesting topic.

\section*{Acknowledgments}

We would like to thank T.G. Tiecke for fruitful discussions and sharing his
trap calculation program. JTMW. thanks J. Dalibard for a valuable discussion.
This work is part of the research program on Quantum Gases of the Stichting
voor Fundamenteel Onderzoek der Materie (FOM), which is financially supported
by the Nederlandse Organisatie voor Wetenschappelijk Onderzoek (NWO).

\appendix

\section{Analytic Model\label{app:Delayed Sudden-Step Model}}

For arbitrary $\phi_{m}$ the position $\boldsymbol{\rho}^{\prime}=\{x^{\prime
},y^{\prime}\}$ with respect to a coordinate system rotated over an angle
$\phi_{m}$ is%
\begin{align}
x^{\prime}(t)  &  =-\rho_{0}\cos\omega t+X_{0}^{\prime}\cos(\Omega
t+\varphi_{x}^{\prime})\\
y^{\prime}(t)  &  =-\rho_{0}\sin\omega t+Y_{0}^{\prime}\sin(\Omega
t+\varphi_{y}^{\prime}),
\end{align}
where $X_{0}^{\prime}$, $Y_{0}^{\prime}$ are the amplitudes and $\varphi
_{x}^{\prime},\varphi_{y}^{\prime}$ the phases with respect to the rotated
axes. Taking the time derivative and using the initial conditions
$\boldsymbol{\rho}^{\prime},\boldsymbol{\dot{\rho}}^{\prime}=0$ at $t=0$,
yields: $\varphi_{x}^{\prime}=\varphi_{y}^{\prime}=0$, $X_{0}^{\prime}%
=\rho_{0}$, $Y_{0}^{\prime}=(\omega/\Omega)\rho_{0}$. This corresponds to an
ellipse with its major axis oriented perpendicular to the instantaneous
direction $\boldsymbol{\hat{\rho}}_{m}^{\prime}\equiv\boldsymbol{\hat{\rho}%
}_{m}$ of the $\mathbf{B}_{m}$ field at $t=0$.

For exponential switch-on of the $\mathbf{B}_{m}$ field with $1/e$ time
$\tau_{0}$, we have for the acceleration in the `primed' coordinate system%
\begin{align}
\ddot{x}^{\prime}(t)  &  =\omega^{2}(1-\exp[-t/\tau_{0}])\rho_{0}\cos\omega
t\ -\nonumber\\
&  \ \ \ \ \ \ \ \ \ \ \ \ \ \ \ \ \ \ \ \ \ \ \ -\Omega^{2}X_{0}^{\prime}%
\cos(\Omega t+\varphi_{x}^{\prime})\\
\ddot{y}^{\prime}(t)  &  =\omega^{2}(1-\exp[-t/\tau_{0}])\rho_{0}\sin\omega
t\ -\nonumber\\
&  \ \ \ \ \ \ \ \ \ \ \ \ \ \ \ \ \ \ \ \ \ \ \ -\Omega^{2}Y_{0}^{\prime}%
\sin(\Omega t+\varphi_{y}^{\prime}),
\end{align}
where, during switch-on, $X_{0}^{\prime}$, $Y_{0}^{\prime}$, $\varphi
_{x}^{\prime}$, $\varphi_{y}^{\prime}$ and $\Omega$ are functions of time.
Since $\Omega\ll\omega$ we may approximate, for $\omega t\ll1$,%
\begin{equation}
\ddot{x}^{\prime}(t)\simeq\omega^{2}(1-\exp[-t/\tau_{0}])\rho_{0}\cos\omega
t\gg\ddot{y}^{\prime}(t). \label{LaunchAcelleration}%
\end{equation}
This shows that the switch-on profile mainly affects the acceleration in the
$x^{\prime}$ direction because this is the initial direction of acceleration.
By the time $\sin\omega t$ is sufficiently large to make $\ddot{y}^{\prime}$
non-negligible, the switch-on transient is already finished. For $\omega
t\ll1$, the velocity in the $x^{\prime}$ direction is given by $\dot
{x}^{\prime}(t)\simeq\omega^{2}(t-\tau+\tau\exp[-t/\tau])\rho_{0}$. This
expression suggests to approximate the switch-on profile by a step function at
$t=\tau_{0}\simeq\tau$ and to treat $X_{0}^{\prime}$, $Y_{0}^{\prime}$,
$\varphi_{x}^{\prime}$, $\varphi_{y}^{\prime}$ as constants for $t\geq\tau
_{0}$. This `delayed sudden-step approximation' is equivalent to imposing the
boundary conditions $\boldsymbol{\rho}^{\prime},\boldsymbol{\dot{\rho}%
}^{\prime}=0$ at $t=\tau_{0}$. In this approximation we obtain for the phases
and amplitudes $\varphi_{x}^{\prime}=\tan^{-1}[\omega^{2}\tau_{0}/\Omega]$,
$\varphi_{y}^{\prime}=0$, $X_{0}^{\prime}=\rho_{0}(1+\omega^{4}\tau_{0}%
^{2}/\Omega^{2})^{1/2}$, $Y_{0}^{\prime}=(\omega/\Omega)\rho_{0}$ for
$t\geq\tau_{0}$. The phase development $\omega\tau_{0}$ due to rotation of the
$\mathbf{B}_{m}$ field during switch-on will appear as a rotation of the major
and minor axes of the macromotion with respect to the primed coordinate system
defined by $t=0$. The optimal value for $\tau_{0}$ can be determined by
comparing the predictions of the analytical model with the results of a
numerical calculation (see Section \ref{sec:Numerical analysis}).

In the presence of gravity, the above analysis remains valid as long as the
radial frequency $\Omega$ does not change substantially during switch-on of
the $\mathbf{B}_{m}$ field; in principle $\Omega$ can be kept constant by
simultaneously switching $B_{0}$ and $B_{m}$ in such a way that the quantity
$\bar{B}_{0}/(1+\frac{1}{2}B_{m}^{2}/\bar{B}_{0}^{2})$ equals the value of
$B_{0}$ before the $\mathbf{B}_{m}$ field was switched on. In case of a small
and fast change in $\Omega$ the above model can be adapted by changing the
initial conditions to $\boldsymbol{\rho}-\{0,\delta y\},\boldsymbol{\dot{\rho
}}=0$ at $t=\tau_{0}$, where $\delta y=\Delta y_{\mathrm{TOP}}-\Delta
y_{\mathrm{IP}}$ is the difference in gravitational sag. Using the adapted
boundary conditions we obtain in the limits $(g\alpha/\bar{B}_{0})^{1/2}%
\ll\Omega\ll\omega$ and $\omega\tau_{0}\ll1$ for the amplitudes and phases of
the macromotion%
\begin{align}
X_{0}  &  =\rho_{0}[1+(\omega^{2}/\Omega^{2}-1)\sin^{2}(\omega\tau_{0}%
+\phi_{m})]^{1/2}\label{X0}\\
Y_{0}  &  =\rho_{0}[1+(\omega^{2}/\Omega^{2}-1)\cos^{2}(\omega\tau_{0}%
+\phi_{m})\nonumber\\
&  \ \ \ \ \ \ \ \ \ \ +\delta y^{2}/\rho_{0}^{2}+2(\delta y/\rho_{0}%
)\sin(\omega\tau_{0}+\phi_{m})]^{1/2}\label{Y0}\\
\varphi_{x}  &  =-\Omega\tau_{0}+\tan^{-1}\left[  (\omega/\Omega)\tan
(\omega\tau_{0}+\phi_{m})\right]  +n\pi\label{Phi-x}\\
\varphi_{y}  &  =-\Omega\tau_{0}+\tan^{-1}[(\Omega/\omega)\{\tan(\omega
\tau_{0}+\phi_{m})\nonumber\\
&  \ \ \ \ \ \ \ \ \ \ \ \ \ \ \ +(\delta y/\rho_{0})/\cos(\omega\tau_{0}%
+\phi_{m})\}]+n\pi. \label{Phi-y}%
\end{align}
where $n=0$ for $|\omega\tau_{0}+\phi_{m}|\leq\pi/2$, $n=1$ for $|\omega
\tau_{0}+\phi_{m}|>\pi/2.$ For $\phi_{m}=0$, these equations coincide with the
equations for $X_{0}^{\prime}$, $Y_{0}^{\prime}$, $\varphi_{x}^{\prime}$ and
$\varphi_{y}^{\prime}$ in the primed coordinate system. Analyzing the limit
$\omega\tau_{0}\rightarrow0$ for the case $\delta y\simeq\rho_{0}$ (typical
for our experimental conditions) we find $X_{0}\simeq\rho_{0}$ and
$Y_{0}\simeq(\omega/\Omega)\rho_{0}$ for $\phi_{m}=0$ and $X_{0}\simeq
(\omega/\Omega)\rho_{0}$ and $Y_{0}\simeq2\rho_{0}$ for $\phi_{m}=\pi/2$.
Thus, we deduce that gravity can have a substantial influence on the amplitude
of the macromotion along its minor axis but not on the amplitude along the
major axis. For known values of $X_{0}$, $Y_{0}$, $\varphi_{x}$ and
$\varphi_{y}$ the angle of rotation $\vartheta$ to align the coordinate system
along the major and minor axis is given by Eq.\thinspace(\ref{eq:rotation}).
For $\Omega\tau_{0}\ll\omega\tau_{0}\ll1$ and in the absence of gravity
$\left(  \delta y=0\right)  $ the delayed sudden-step gives rise to a small
rotation $\Delta\varphi\simeq\omega\tau_{0}$, independent of $\phi_{m}$. For
$\delta y\simeq\rho_{0}$ gravity gives rise to an additional contribution to
this rotation, which is minimal for $\phi_{m}=\pi/2$, where the macromotion is
launched perpendicular to the gravity direction and Eq.\thinspace
(\ref{eq:rotation}) can be approximated by $\varphi\simeq\omega\tau
_{0}[1+\left(  1+\delta y/\rho_{0}\right)  (\Omega/\omega)^{2}]$. The
contribution is maximal for $\phi_{m}=0$, where Eq.\thinspace
(\ref{eq:rotation}) can be approximated by $\varphi\simeq\omega\tau
_{0}-(\Omega/\omega)^{2}\left(  \delta y/\rho_{0}\right)  ^{2}(1+\omega
\tau_{0})$.

Insight in the dependence of $X_{0}$ and $Y_{0}$ on the settling behavior of
$B_{0}$ can be obtained from Eq.\thinspace(\ref{LaunchAcelleration}), which
shows that the initial acceleration and, hence, the launch speed scales with
the initial value of $\rho_{0}$. Therefore, most of the settling behavior is
captured by using the \textit{initial} value of $\rho_{0}$ in Eqs.\thinspace
(\ref{X0}) and (\ref{Y0}). The $5\%$ change in $\Omega$ requires a further
refinement of the model. This cannot be implemented without sacrificing the
simplicity of the model and is not pursued here.


\begin{thebibliography}{99}
\bibitem {ApproxPhase}We verified that the approximation $\phi=\phi_{m}+\pi$
holds to within one degree for conditions relevant for our analysis.
\end{thebibliography}
\end{document}